\documentstyle[12pt,epsfig]{article}
\newcommand{\blankline}{\vskip .3cm}

\newcommand{\coe}{{\bar{\epsilon}}}

\newcommand{\beq}{\begin{equation}}
\newcommand{\eeq}{\end{equation}}
\newcommand\bea{\begin{eqnarray}}
\newcommand\eea{\end{eqnarray}}

\newcommand{\id}{\mbox{id}}\newcommand{\sre}{{S_R{}^E}}
\newcommand{\dimn}{\mbox{dim}}

\makeatletter
\makeatletter
\newcommand{\ps@preprint}
\makeatother
\begin{document}
\centerline{\LARGE Coarse graining in spin foam models} 
\blankline
\blankline \centerline{Fotini Markopoulou${}^*$} 
\blankline
\centerline{\it Perimeter Institute for Theoretical Physics}
\centerline{\it 35 King Street North, Waterloo, Ontario N2J 2W9,
Canada} 
\centerline{\it and} 
\centerline {\it Department of
Physics, University of Waterloo} 
\centerline{\it Waterloo, Ontario
N2L 3G1, Canada} 
\blankline \blankline 
\centerline{March 11, 2002} 
\vfill
\centerline{\bf Abstract} 
\blankline
We formulate the problem of
finding the low-energy limit of spin foam models as a
coarse-graining problem in the sense of statistical physics.  This
suggests that renormalization group methods may be used to find
that limit.  However, since spin foams are models of spacetime at
Planck scale, novel issues arise: these microscopic models are
sums over irregular, background-independent lattices. We show
that all of these issues can be addressed
by the recent application of the Kreimer Hopf algebra for quantum
field theory renormalization to non-perturbative statistical
physics. The main difference from standard renormalization group
is that the Hopf algebra executes block transformations in
{\em parts} of the lattice only but in a controlled manner so that
the end result is a fully block-transformed lattice. 
\vfill
${}^*$fotini@perimeterinstitute.ca 
\eject 
\tableofcontents 
\eject

\section{Introduction}

Progress on several fronts in the search for a quantum theory of
gravity has resulted in the first detailed models of the
microscopic structure of spacetime: spin foams.  The first such
models \cite{Rei1,Rei2,ReRo,MaSm1,Bae1} were based on the predictions for
Planck scale geometry from loop quantum gravity and, in
particular, the discovery that spatial areas and volumes have
discrete spectra \cite{RoSm} (for a recent thorough review of loop
quantum gravity, see \cite{Thie}).  Currently, there exist several
such models, which also include structures from other approaches
to quantum gravity, from Lorentzian path integrals 
\cite{AmLo,BaCrL,PeRoL} and
causal sets \cite{Ma1,MaSm2} to euclidean general
relativity \cite{Rei2,BaCrE,Iwas,PeRo}, 
and from string networks \cite{MaSm3} to
topological quantum field theories (see \cite{Bae2}). For reviews
see \cite{Bae2,Orit}.

In spin foam models, the microscopic degrees of freedom are
representations and intertwiners of the appropriate group
(originally SU(2)) and live on a branched 2-surface, or 2-complex.
A specific model is given by a partition function that sums over
all microscopic degrees of freedom and all (model-dependent)
weights on the vertices of the 2-complex. It also sums over all
2-complexes that interpolate between the given in and out
3-geometry states, making spin foams a path-integral approach to
quantum gravity.

Clearly, a spin foam model will be a good candidate for a quantum
theory of gravity only if it can be shown to have a good low
energy limit, which contains the known theories, namely, general
relativity and quantum field theory. As in \cite{Ma4}, in this
paper we suggest that this should be treated as a problem in
statistical physics.  That is, for spin foams in the correct class
of  microscopic models, we should find the known macroscopic
theory by integrating out the microscopic degrees of freedom. What
is required, then, is a renormalization group approach for spin
foams.  This appears promising since spin foams are given by a
partition function which has the same functional form as that for
a spin system or lattice gauge theory (apart from the sum over all
2-complexes). Also, spin foam models are ``atomic'', that is,
there is a minimum length, the Planck scale.

The most intuitive way of understanding the renormalization group
in statistical physics is 
real space renormalization, where we coarse-grain the lattice.  The
aim of this paper is to set up such a coarse-graining method for
spin foams.  We can list the basic features of the problem of spin
foam coarse-graining:
\begin{enumerate}
\item
The microscopic degrees of freedom are representations of a group
or algebra.  A block transformation on a spin foam involves (exact
or approximate) summing over the possible values these can take.
This is cumbersome, but conceptually straightforward.  (Such
calculations have been carried out, for example, in \cite{PeRo}.)
\item
The weights in the partition function are amplitudes rather than
probabilities.
\item
The lattices are the highly irregular spin foam 2-complexes.
Simplifying the problem by restricting to regular 2-complexes does
not appear to be an option.  It would be a drastic modification of
the model, as these are rare configurations in the full sum.
\item
Spin foams are background-independent.  This means that we cannot
use a global lattice spacing as a parameter in our coarse-graining
procedure. However, there is a minimum length, the Planck length
$l_{\rm Pl}$, so it is clear what a {\em single} block
transformation does: it increases the scale from $l_{\rm Pl}$ to a
multiple of it.  This is equivalent to increasing the lattice
spacing, but the problem is that, without a background and on an
irregular lattice, we cannot do this everywhere at the same time,
as is required to obtain a configuration space renormalization
group equation.
\item
A spin foam partition function has the same functional form as a
lattice gauge theory or a spin system, except for one important
difference: the spin foam partition function contains a sum over
{\em all} 2-complexes with the given boundary. Therefore, we need
to coarse-grain {\em sums} over lattices.
\end{enumerate}

1, 2 and 3 above are technically challenging, but there are
existing examples of such problems in statistical physics. 4 and
5, however, are novel issues, due to the fact that these are
microscopic models of spacetime itself.

In this paper, we address 4 and 5 and propose a method to deal
with them.  Essentially, we generalize the renormalization group
transformation for a lattice gauge system to a set of local block
transformations and this directly addresses 4 (by ``local'' we
mean that we block transform parts of the lattice only). These
local transformations are equivalence relations in a Hopf algebra,
thus the order in which they are performed is controlled.
Furthermore, the elements of the algebra {\em are} sums over spin
foams, so 5 is directly taken into account.  It is not surprising
that this method of coarse-graining differs from standard
renormalization group. Roughly speaking, the usual renormalization
group is embedded in the Hopf algebra.

The outline of this paper is as follows.  In the next Section we
briefly review the definition of spin foams and the sense in which
they are microscopic models of spacetime.  For coarse-graining
purposes (configuration space renormalization group), spin foams, like
any other lattice model, have to be partitioned into the appropriate
subfoams that will be block-transformed.  In Section 3, we define
labelled partitioned spin foams and subfoams, and the corresponding
(parenthesized) weights.  These are the elements of the Hopf algebra
whose operations are given in Section 4.  Again, the labelled
partitioned spin foams correspond to parenthesized weights, and the
Hopf algebra operations of such weights is given in Section 5.  A
renormalization group transformation is an equivalence relation in
this algebra (Section 6).  By incorporating this equivalence relation
into the antipode, in Section 7, we use this modified antipode to
perform local scale transformations.  This is essentially a
generalization of the renormalization group equation, and we
illustrate this in Section 8 by using the modified antipode to
coarse-grain an ordinary regular square lattice.  In Section 9, we
discuss the general form of the modified antipode.  We discuss our
results and the possibilities for applying our method to specific spin
foam models in the Conclusions.

The Hopf algebra we use is the one that Kreimer  showed underlies
the renormalization of quantum field theory \cite{Kr1,Kr2}.  In
general, the elements of this algebra are rooted trees. In quantum
field theory, the branches represent subdivergencies of a Feynman
diagram. In our case, the branches are spin subfoams. In Kreimer's
case, the Hopf algebra takes care of the combinatorial part of the
problem of renormalization in quantum field theory. However, its
structure is very general and it underlies block transformations
in the non-perturbative renormalization group, as we showed in
\cite{Ma4}.  We can transcribe Kreimer's algebra to spin foams
even though the physics that dresses the rooted trees is very
different from QFT:  in spin foams we do have a UV cutoff and want
the large $N$ limit.  

Closing the Introduction, we would like to stress that, as in other
problems in statistical physics, the renormalization group and its
variation that we propose here, is more like a general set of ideas
than a recipe which can be directly applied to any system.  Further
progress should be made by analyzing individual models as well as by
experimental input.

\section{Spin foams as models of the microscopic structure of spacetime}

A spin foam is a labelled 2-complex whose faces are labelled by
representations of some group $G$, the edges  by intertwiners in
the group, and the vertices carry the evolution amplitudes.  These
are functions of the faces and edges that meet on that vertex that
code the evolution rules for the model (figure 1).

\begin{figure}
\center{
\epsfig{file=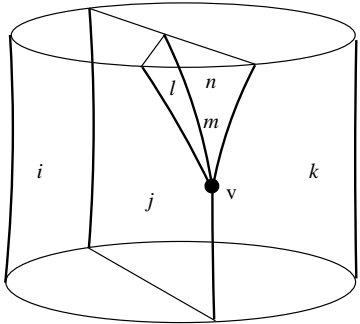}}
\caption{A spin foam}
\end{figure}

A spin foam model is given by a partition function of the form
\beq
Z(s_{i},s_{f})
=\sum_{\Gamma} N(\Gamma) \sum_{{\mbox{\footnotesize labels on
}}\Gamma} \prod_{f\in\Gamma} \dimn j_{f} \prod_{v\in\Gamma}
A_{v}(j).
\label{eq:Z}
\eeq
The first sum is over all spin foams
$\Gamma$ interpolating between a given initial spin network\footnote{
A spin network is a graph whose edges are labelled by representations
of the group $G$ and its nodes are labelled by the intertwiners.
Here, we regard spin networks as the
``spacelike slices''  of a spin foam:  If we take a
spacelike cut through a spin foam, we obtain a graph.  Its edges are
cuts through the spin foam faces, and so we label them with the
same representations.  Its nodes are cuts of the
spin foam edges, and we label them with intertwiners.
{\bf Notation:} In the literature, ``spin network'' often
refers to a graph labelled by $SU(2)$
representations.  For us, spin networks are the spacelike cuts
through spin foams, and so we use the name for graphs carrying
generic $G$ representations.  }
$s_{i}$ and a final one $s_{f}$.  $\dimn j_{f}$ is
the dimension of the $G$ representation labelling the face $f$ of
$\Gamma$. $A_{v}$ is the amplitude on the vertex $v$ of $\Gamma$,
a given function of the labels on the faces and edges adjacent
to $v$. A choice of the group $G$ and the functions $A_{v}$ (and
possibly a restriction on the allowed 2-complexes) defines a
particular spin foam model.  For reviews of spin foams see
\cite{Bae2,Orit}.

Much of the motivation for considering spin foam models comes from
the surprising result of loop quantum gravity:  the quantized
spatial area and volume are discrete \cite{RoSm}.
$SU(2)$ spin networks are the basis
states of the loop quantization of general relativity.
They diagonalize the quantum area and volume
operators whose spectrum was discovered to be discrete
(see \cite{Thie} and references therein).

In loop quantum gravity, the spin networks are embedded in the
spatial manifold of the classical canonical theory.
Here we are considering abstract spin foams, that is, spin
foams which are combinatorial cell complexes and not embedded in some
preexisting continuous manifold.
Spacelike cuts through abstract $SU(2)$ spin foams are Penrose
spin networks, the combinatorial graphs Penrose used to recover angles
in 3-space from the rules of angular momentum \cite{Penr}.

As mentioned in the introduction, other degrees of freedom such as
matter or supersymmetric ones can be introduced by using or adding
the appropriate group representations.

Currently, there are several spin foam models, candidates for the
microscopic structure of spacetime.  The very first test such a
model has to pass is to have a good low energy limit, where it
reproduces the known theories, general relativity and quantum field
theory.  What that limit means is not yet clear.  Our working assumption
here is that spin foams are models of spacetime similar to the way that
spin systems are microscopic models of matter.  What is then required
is that, in the low-energy limit, the spin foam observables, 
coarse-grained
over many Planck lengths, agree with those in a classical spacetime.

In solid state physics, the correspondence between the microscopic
spin system model and the macroscopic matter is obtained using
renormalization group techniques.  We propose that the same basic idea 
should
also apply to spin foams.
The starting point in this paper is the observation that in equation
(\ref{eq:Z}), $Z_{\Gamma}$ for a single spin foam is a generalization
of the partition function of a spin system or lattice gauge system to
variables that are group representations and functions of these
representations.  Since $Z_{\Gamma}$ factorizes to a product over
local labels and functions, it is possible to carry out block
transformations by summing over internal variables.

In the next Section, we describe what block transformations involve for
a spin foam.


\section{Block transformations and partitioned spin foams}

A renormalization group transformation of a lattice $\Gamma$
is a procedure that involves two steps.  One is the calculation of a
typical block transformation in the theory.  This involves picking
some small sublattice $\gamma$ of $\Gamma$ that carries only few
microscopic degrees of freedom.  A block transformation eliminates
some of them, either by summing over all their possible values, or by
some approximate recipe (decimation, truncation etc).

For a spin foam theory, an example of such a block transformation is
the calculation by Perez and Rovelli on the modified
Barrett-Crane model \cite{PeRo}.
This is a block transformation that sums over all internal variables.
For example, the spin foam
\[
    \gamma=\begin{array}{c}\mbox{\epsfig{file=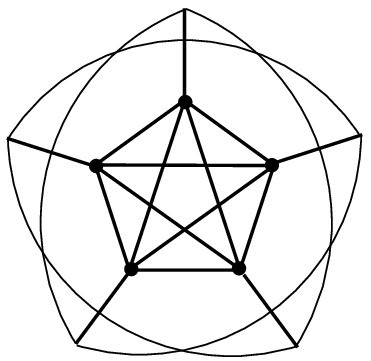}}\end{array}
\]
is replaced
by the coarse-grained spin foam
\[
    \gamma'=\begin{array}{c}\mbox{\epsfig{file=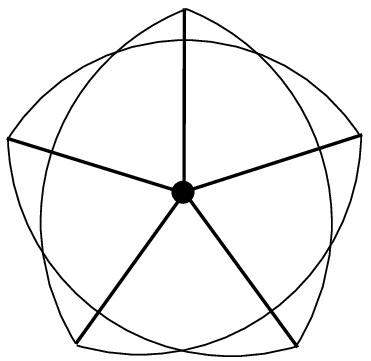}}\end{array}
\]
where a summation has been performed over all allowed values of the 
labels
on the faces we eliminated.

Having obtained the functional form of such a block transformation,
the next step is to repeatedly apply it on the entire spin foam
$\Gamma$ to obtain a coarse-grained one, $\Gamma'$.

In lattice gauge theory, or in spin systems,
the lattices are usually regular and
this is an unambiguous procedure.  We
simply partition the lattices uniformly into sublattices of the form we 
have
used for the calculation of the single block transformation, and
perform this transformation on each such sublattice.

In the spin foam case, however, the 2-complexes are highly
irregular.  This makes this second step a non-trivial combinatorial
problem.  The algebraic method we propose in this article deals with this
second part of a renormalization group transformation (and also deals 
with
the issue of the choice of a scaling parameter in a background 
independent
theory, as we will discuss in Section 7).

Our motivation is a rather simple observation which we will motivate
diagrammatically on a 2-dimensional spin foam.
Suppose I have calculated a block transformation
\[
\gamma =\begin{array}{c}\mbox{\epsfig{file=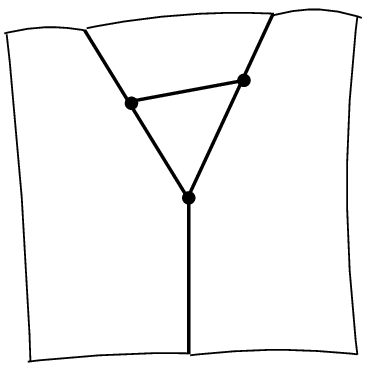}}\end{array}
\qquad\longrightarrow \qquad
\gamma'=\begin{array}{c}\mbox{\epsfig{file=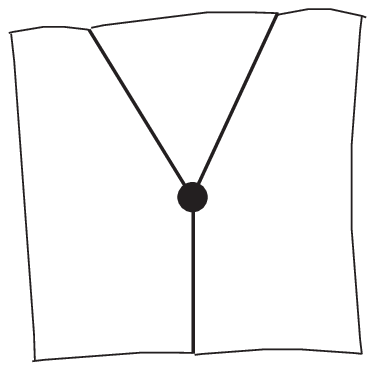}}\end{array}
\]
(the fat node denotes a renormalized vertex),
and now I wish to block-transform the larger lattice
\[
\Gamma =\begin{array}{c}\mbox{\epsfig{file=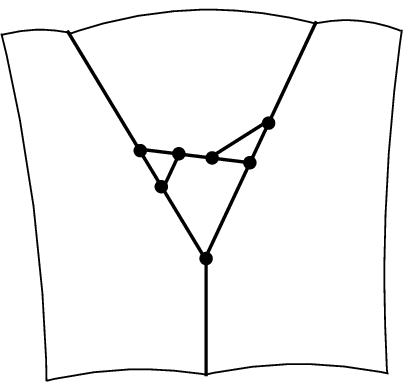}}\end{array}.
\]
To do so, I should first choose a {\em partitioning} of $\Gamma$ into
subfoams, such that the smallest ones are of the same form as
$\gamma$.  Ideally, I should choose a partition such that,
when I have block-transformed away
all the smallest subfoams, I will be left with next-smallest
subfoams which again
have the same form as $\gamma$, so I can use again the same block
transformation. Such a partition is, for example,
\beq
\Gamma=\begin{array}{c}\mbox{\epsfig{file=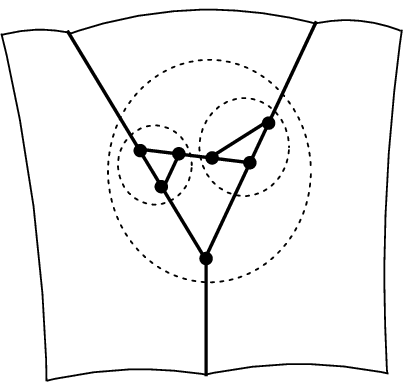}}\end{array}.
\eeq
Block-transforming $\Gamma$ according to the marked partition,
gives the sequence
\[
\begin{array}{c}\mbox{\epsfig{file=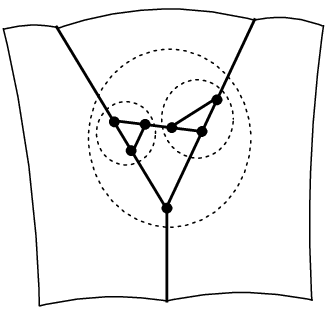}}\end{array}
\longrightarrow
\begin{array}{c}\mbox{\epsfig{file=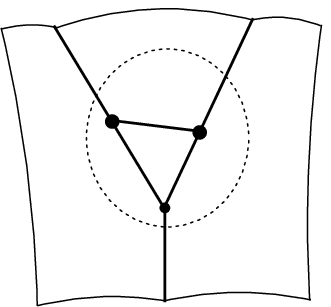}}\end{array}
\longrightarrow
\begin{array}{c}\mbox{\epsfig{file=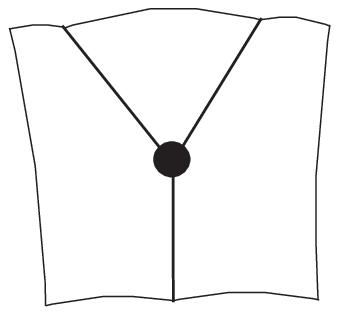}}\end{array}.
\]

It is this nesting structure of subfoams
that we will use in this paper to construct an algebra for
coarse-graining.   As we saw, to begin coarse-graining a lattice, we
need to first partition it into sublattices.  It is not surprising,
therefore, that the elements of our algebra will be {\em partitioned
spin foams} and their corresponding weights.   In the remainder of
this section we define partitioned spin foams and parenthesized spin
foam weights.

\subsection{Partitioned spin foam 2-complexes}

\begin{figure}
\center{
\epsfig{file=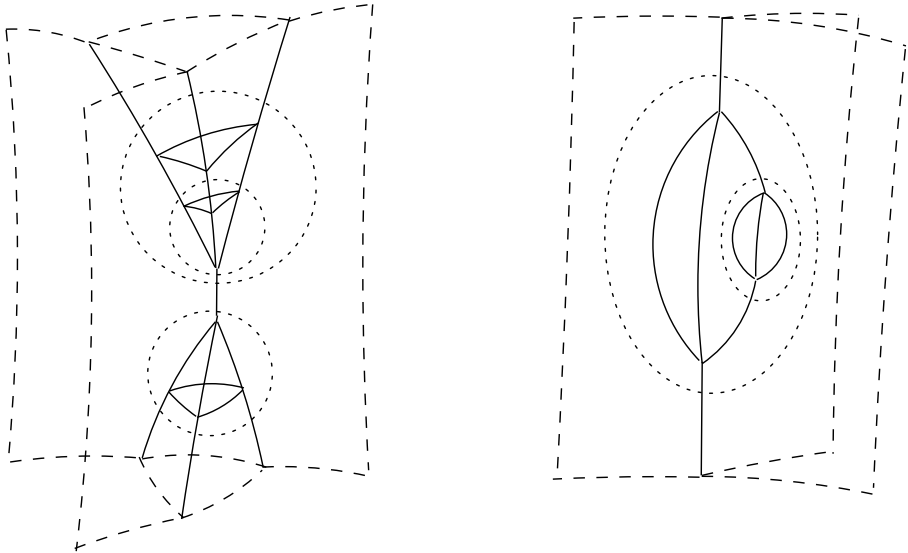}}
\caption{Two examples of subfoams.}
\end{figure}

A {\em subfoam} $\gamma$ of a spin foam $\Gamma$ is a subset of the
faces of $\Gamma$, together with any vertices and edges that are
boundaries of these faces:
\[
\Gamma=\begin{array}{c}\mbox{\epsfig{file=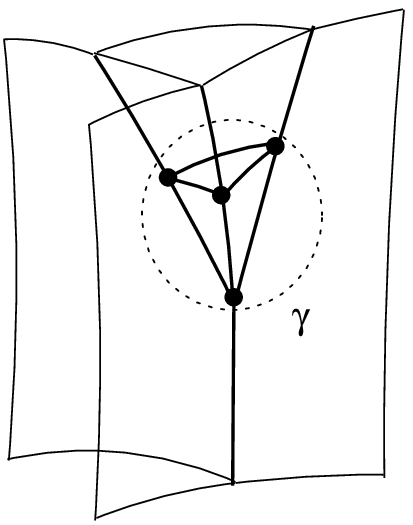}}\end{array}
\quad\longrightarrow\quad
\gamma=\begin{array}{c}\mbox{\epsfig{file=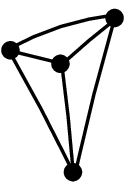}}\end{array}.
\]
The weight of a subfoam $\gamma$ is
\beq
\omega_\gamma=\prod_{f\in\gamma}\dimn j_f
    \prod_{v\in\gamma} A_v.
\eeq

A subfoam is {\em connected} if every face in $\gamma$ shares at
least one edge with some other face in $\gamma$.
A subfoam may be a set of disconnected subfoams.

A subfoam $\gamma_1$ is {\em nested} in $\gamma_2$,
$\gamma_1\subset\gamma_2$, if the set of faces of $\gamma_1$ is a
proper subset of the faces of $\gamma_2$.  Two subfoams are
{\em disjoint}, $\gamma_1\cap\gamma_2=\emptyset$, if they have
no faces, edges or vertices in common.\footnote{
The edges of spin foams are labelled by intertwiners of the
group $G$.  However, by appropriate normalizations,
these can be absorbed in the vertex amplitudes.
We have not, therefore, included labels on the edges in the
partition function, and we will not worry about them
in the definition of an allowed partition.}
When two subfoams are neither nested nor disjoint, they are
{\it overlapping}.

A {\em partitioned spin foam} is a spin foam 2-complex $\Gamma$ marked 
with
an {\em allowed} partition into subfoams $\{\gamma_{i}\}$.
``Allowed'' means that
there are no overlapping subfoams, that is, any two subfoams
$(\gamma_{1}, \gamma_{2})$ in the partition are either nested:
$\gamma_{1}\subset\gamma_{2}$ or $\gamma_{2}\subset \gamma_{1}$, or
they are disjoint: $\gamma_{1}\cap\gamma_{2}=\emptyset$.

The {\em remainder} $\Gamma/\gamma$ of the subfoam $\gamma$ is the
2-complex we obtain by {\em deleting} $\gamma$ from $\Gamma$, that
is, eliminating all faces of $\gamma$ and joining the vertices on
the boundary of
each of its connected components into a single vertex for that
component.
There is {\em no} amplitude on that vertex and we
will denote this in diagrams by a white vertex:
\[
\Gamma=\begin{array}{c}\mbox{\epsfig{file=part.eps}}\end{array}
\quad\longrightarrow\quad
\Gamma/\gamma=\begin{array}{c}\mbox{\epsfig{file=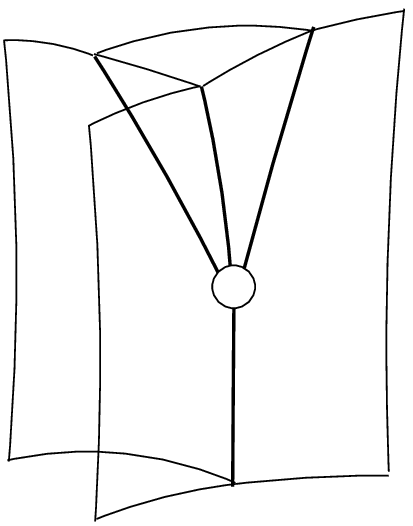}}\end{array}
\]
Naturally, the remainder is also a partitioned spin foam, with the 
partition it
inherits from $\Gamma$.
 {\em In the rest of this paper we will only work with partitioned spin
foams, but we will call them simply spin foams. }

While the spin foam 2-complex is a useful pictorial representation of
a contribution in the partition function (\ref{eq:Z}) of the theory,
the physical content lies in the corresponding weight.  Next, we
consider the analogue of the nesting structure of spin foams for
their weights.

\subsection{Parenthesized spin foam weights}
\label{weights}

Given a  spin foam $\Gamma$, the weight it contributes to $Z$ is
$\omega_{\Gamma}=\prod_{f}\dimn j_{f}\prod_{v} A_{v}$ in the notation
defined in section 2.  This is the physical content of the spin foam,
and the quantity we wish to coarse-grain.

We will represent the weight of a partitioned spin foam by a
{\em parenthesized weight}.  That is, we enclose in brackets the
factors in $Z$ that correspond to the subfoams in the partition.  For
nested subfoams, we obtain nested brackets, $(())$, and for disjoint
subfoams disjoint brackets, $()()$.  For example, the parenthesized
weight for the 2-dimensional spin foam
\beq
\Gamma= \begin{array}{c}
\mbox{\epsfig{file=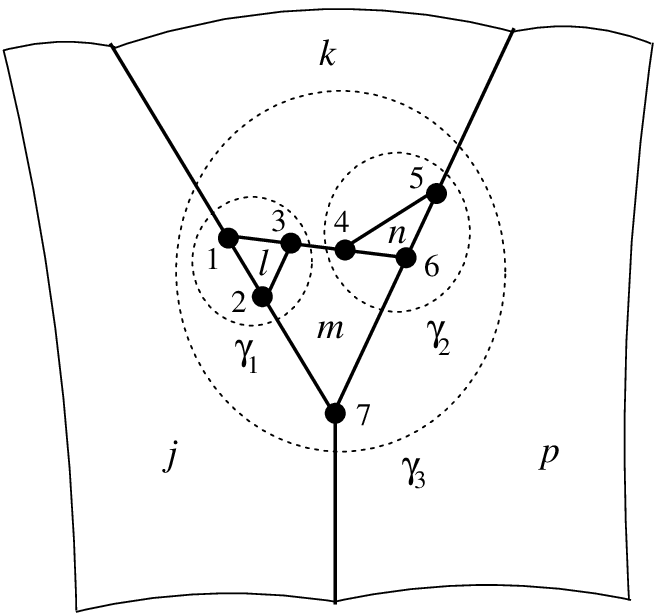}}
\end{array}
\eeq
is
\bea
\omega_{\Gamma}&=&\left(\left(\left(\omega_{\gamma_{1}}\right)
\left(\omega_{\gamma_{2}}\right)
\omega_{{\gamma_3}/{\gamma_1\cup\gamma_2}}\right)
\omega_{\Gamma/{\gamma_3}}
\right) \cr
&=&
\left(\left(\left(d_l A_{v_{1}}A_{v_{2}}A_{v_{3}}\right)
\left( d_n A_{v_{4}}A_{v_{5}}A_{v_{6}}\right)
d_m  A_{v_{7}}\right)
d_j d_k d_p\right).
\eea
where we have used $d_i$ as shorthand for dim$\/i$.

An example in 3 dimensions is the spin foam
\[
\Gamma= \begin{array}{c}
\mbox{\epsfig{file=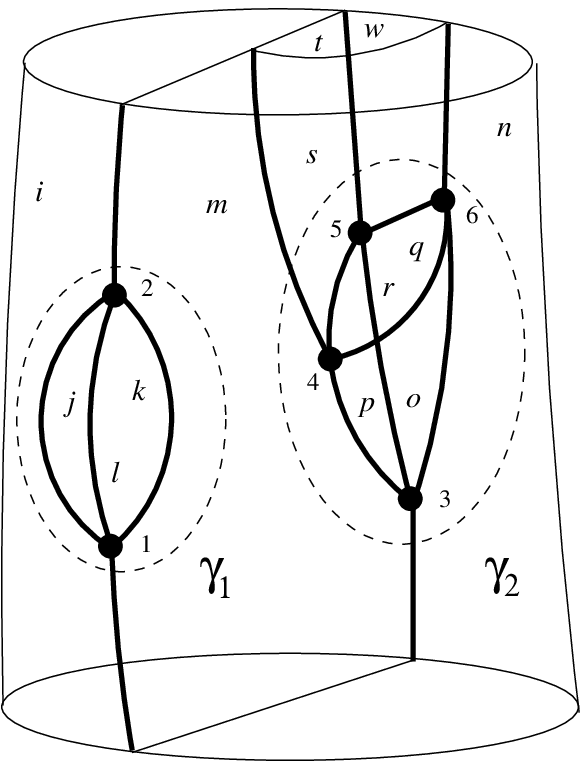}}
\end{array}
\]
This has weight
\beq
\begin{array}{rl}
\omega_\Gamma =& \left(\left(
    \omega_{\gamma_1}\right)\left(\omega_{\gamma_2}\right)
    \omega_{\Gamma/{\gamma_1\cup\gamma_2}}\right)\\
    =& \left(\left(
    d_i d_k d_l A_1 A_2 \right)
    \left( d_o d_p d_q d_r A_3 A_4 A_5 A_6 \right)
      d_i d_m d_n d_s d_t d_w \right).
\end{array}
\eeq


\section{The Hopf algebra of partitioned spin foams}

Partitioned spin foams form a Hopf algebra whose operations we list in
this section.  Let $V$ be the vector space of spin foams over the
complex numbers ${\bf C}$.  A general element of $V$ has the form
$\Gamma=\sum_{i} c_{i}\Gamma_{i}$, $c_{i}\in{\bf C}$.  Let $e$ be the
{\em empty spin foam}.  The following operations can be defined on
$V$:

$\bullet$ {\bf Multiplication.} Multiplication
 $m:V\otimes V\rightarrow V$ is the disjoint union of spin foams:
\beq
m(\Gamma_{1}\otimes\Gamma_{2})=\Gamma_{1}\cdot\Gamma_{2}:=
\Gamma_{1}\cup\Gamma_{2}.
\eeq
Since the order of multiplying the weights of two disjoint
spin foams in $Z$ does not
matter, multiplication is commutative.
Further, $\Gamma\cdot e =e\cdot\Gamma=\Gamma$ for all $\Gamma\in V$.

$\bullet$ {\bf Unit.}
The {\em unit} operation $\epsilon:{\bf C}\rightarrow V$ creates spin
foams:
\beq
\epsilon(c)=ce\qquad c\in{\bf C}.
\eeq

$\bullet$ {\bf Counit.}  The {\em counit} $\coe:V\rightarrow {\bf C}$
annihilates all non-empty spin foams:
\beq
\coe(\Gamma)=\left\{ \begin{array}{lcr}
       0& &\Gamma\neq e\\
       1& &\Gamma=e
       \end{array}
       \right. .
\eeq

$\bullet$ {\bf Coproduct.}
The partitioning of a spin foam lets us define an
operation that unfolds the nesting structure of its subfoams.
This is the {\em coproduct} $\Delta:V\rightarrow V\otimes V$,
which splits a spin foam $\Gamma$ into all possible pairs of
subfoams  in the given partition paired with their remainders:
\beq
\Delta(\Gamma)=\Gamma\otimes e +e\otimes\Gamma
+\sum_{\gamma}\gamma\otimes{\Gamma/\gamma}.
\label{eq:coproduct}
\eeq
The subfoams $\gamma$ in the above sum range over
all subfoams in the partition of $\Gamma$,
except the empty one and $\Gamma$ itself which we wrote separately as
the first two terms.

On the empty spin foam $e$, we get $\Delta(e)=e\otimes e$
and, for a product we have $\Delta(\Gamma_1\cdot\Gamma_2)=
\Delta(\Gamma_1)\Delta(\Gamma_2)$.
Also, if $\gamma_p$ is a spin foam that has no subfoams,
\beq
\Delta(\gamma_p)=\gamma_p\otimes e+e\otimes\gamma_p.
\eeq
We call such a spin foam {\em primitive}.  One can check
that $\Delta$ is coassociative but not cocommutative\footnote{
Partitioned spin foams as well as their parenthesized
weights are labelled rooted trees as in Kreimer \cite{Kr1}.   Therefore, 
for
coassociativity, cocommutativity, properties of the antipode etc,
we refer the reader to \cite{Kr1,Kr2}.}.

The coproduct is a very important operation for our coarse-graining.
Essentially, it identifies the subfoams that we should block
transform (those that appear on the left of the tensor products) and
so it prepares the spin foam for coarse-graining.

\begin{quote}
{\small
{\bf Example 1.}  The coproduct for the spin foam
\beq
\Gamma=\begin{array}{c}\mbox{\epsfig{file=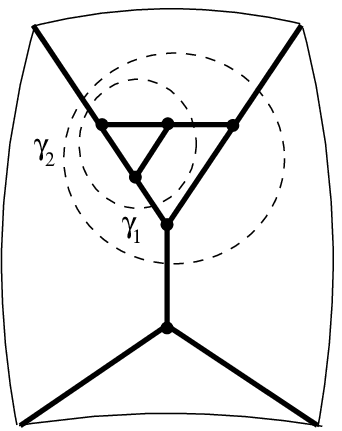}}\end{array}
\label{eq:example}
\eeq
with the marked partition is
\beq
\begin{array}{rl}
\Delta(\Gamma)&=\Gamma\otimes e +e\otimes\Gamma
    +\gamma_1\otimes \Gamma/\gamma_1
    +\gamma_2\otimes \Gamma/\gamma_2\\
    &=
    \begin{array}{c}\mbox{\epsfig{file=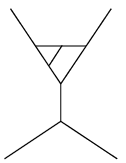}}\end{array}\otimes e
    +e\otimes \begin{array}{c}\mbox{\epsfig{file=f7.eps}}\end{array}
    +\begin{array}{c}\mbox{\epsfig{file=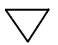}}\end{array}
    \otimes
    \begin{array}{c}\mbox{\epsfig{file=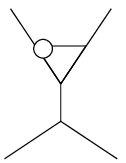}}\end{array}\\
    &\ +\begin{array}{c}\mbox{\epsfig{file=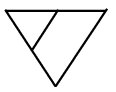}}\end{array}
    \otimes
    \begin{array}{c}\mbox{\epsfig{file=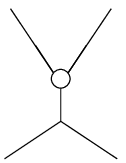}}\end{array}.
    \end{array}
\eeq

{\bf Example 2.}  The spin foam
\beq
\Gamma=\begin{array}{c}\mbox{\epsfig{file=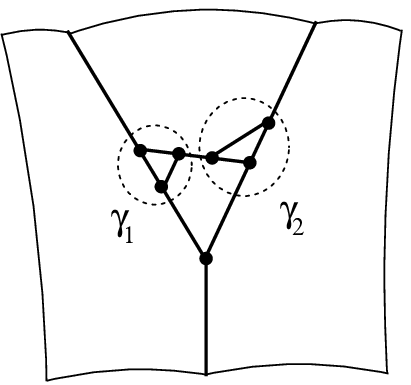}}\end{array}
\label{eq:example2}
\eeq
containing two disjoint spin foams $\gamma_{1}$ and $\gamma_{2}$ has
coproduct
\beq
\begin{array}{rl}
\Delta(\Gamma)=&\Gamma\otimes e+ e\otimes\Gamma
+\gamma_{1}\otimes\Gamma/\gamma_{1}
+\gamma_{2}\otimes\Gamma/\gamma_{2}
+\gamma_{1}\gamma_{2}\otimes\Gamma/{\gamma_{1}\gamma_{2}}\\
=&\begin{array}{c}\mbox{\epsfig{file=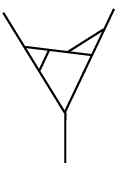}}\end{array}\otimes e
+
e\otimes\begin{array}{c}\mbox{\epsfig{file=ex20.eps}}\end{array}
+\begin{array}{c}\mbox{\epsfig{file=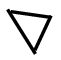}}\end{array}
\otimes
\begin{array}{c}\mbox{\epsfig{file=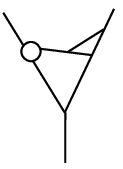}}\end{array}\\
&+
\begin{array}{c}\mbox{\epsfig{file=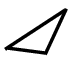}}\end{array}
\otimes
\begin{array}{c}\mbox{\epsfig{file=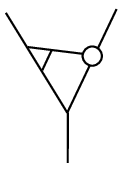}}\end{array}
+
\begin{array}{c}\mbox{\epsfig{file=ex23.eps}}\end{array}
\begin{array}{c}\mbox{\epsfig{file=ex24.eps}}\end{array}
\otimes
\begin{array}{c}\mbox{\epsfig{file=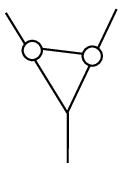}}\end{array}.
\end{array}
\eeq

}
\end{quote}

$\bullet$ {\bf Antipode.}
Can we define the {\em inverse} of a spin foam?  It should
be an operation ${\mbox \em Inv}:V\rightarrow V$, with the property
\beq
{\mbox \em Inv}(\Gamma) \cdot\Gamma =
\Gamma\cdot{\mbox \em Inv}(\Gamma)=1
\label{eq:inv}
\eeq
for all $\Gamma\in V$.

There is no natural definition of the inverse of a
spin foam.  However, we can define a generalization of the inverse.
First, let us rewrite (the left half of) the defining equation
(\ref{eq:inv}) of the inverse as
\beq
{\mbox \em Inv}(\Gamma)\cdot\Gamma \equiv m\left({\mbox \em
Inv}\otimes\id\right)\left(\Gamma\otimes\Gamma\right)=1 .
\eeq
We now replace $(\Gamma\otimes\Gamma)$
with the coproduct on $\Gamma$, and 1 with our map to the complex
numbers, $\coe(\Gamma)$.  Finally, let us name $S$ the generalization of 
${\mbox
\em Inv}$.  So we have
\beq
m\left(S\otimes\id\right)\left(\Delta(\Gamma)\right)=\coe(\Gamma).
\label{eq:O}
\eeq
An operation $S:V\rightarrow V$ satisfying the above equation, and
also
\beq
m(\id\otimes S)\left(\Delta(\Gamma)\right)=\coe(\Gamma),
\label{eq:Or}
\eeq
is called an {\em antipode}.

Equations (\ref{eq:O}) and (\ref{eq:Or}) are the defining property of the 
antipode.  We
still need to give an explicit expression for the action of $S$ on a
partitioned spin foam.  It is the iteration
\beq
\begin{array}{rl}
S(\Gamma)&=-\Gamma-\sum_{\gamma}S(\gamma)\cdot\Gamma/\gamma\\
S(\gamma_{p})&=-\gamma_{p},
\end{array}
\label{eq:S}
\eeq
which stops when a primitive lattice $\gamma_p$ is reached in the sum.
As before, the subfoams in the sum range over all proper subfoams in
the given partition of $\Gamma$ (i.e.\ excludes $\Gamma$ and $e$).

On a product of spin foams we have
$S(\Gamma_1\cdot\Gamma_2)=S(\Gamma_1)\cdot S(\Gamma_2)$,
while $S(e)=e$.

It is straightforward to check, using (\ref{eq:coproduct}), that $S$
as defined in (\ref{eq:S}) satisfies (\ref{eq:O}) and (\ref{eq:Or})
for all $\Gamma\in V$ and therefore it is an antipode for $V$.
One can also check that $S^2=\id$.  Note that this depends on the
commutativity of the product.

\begin{quote}
{\small
{\bf Example 1.}
The action of $S$ on the example spin foam (\ref{eq:example}) gives
\beq
S(\Gamma)=-\Gamma-S(\gamma_1)\cdot\Gamma/\gamma_1
    -S(\gamma_2)\cdot\Gamma/\gamma_2.
\label{eq:Sexample}
\eeq
The subfoam $\gamma_1$ is primitive and therefore
\beq
S(\gamma_1)=-\gamma_1
    =-\begin{array}{c}\mbox{\epsfig{file=f1.eps}}\end{array}.
\label{eq:Sexample2}
\eeq
The subfoam $\gamma_2$ gives
\beq
\begin{array}{rl}
S(\gamma_2)&=-\gamma_2-S(\gamma_1)\cdot\gamma_2/\gamma_1\\
    &=-\begin{array}{c}\mbox{\epsfig{file=f2.eps}}\end{array}
    +\begin{array}{c}\mbox{\epsfig{file=f1.eps}}\end{array}
    \begin{array}{c}\mbox{\epsfig{file=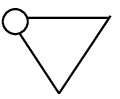}}\end{array}.
\end{array}
\label{eq:Sexample3}
\eeq
Inserting  (\ref{eq:Sexample2}) and (\ref{eq:Sexample3}) in
(\ref{eq:Sexample}), we find that the antipode of $\Gamma$
is
\beq
\begin{array}{rl}
S(\Gamma)&=-\Gamma+\gamma_1\cdot \Gamma/\gamma_1
    +\gamma_2\cdot\Gamma/\gamma_1
    -\gamma_1\cdot\gamma_2/\gamma_1\cdot
        \Gamma/\gamma_2\\
    &=-\begin{array}{c}\mbox{\epsfig{file=f7.eps}}\end{array}
    +\begin{array}{c}\mbox{\epsfig{file=f1.eps}}\end{array}
    \begin{array}{c}\mbox{\epsfig{file=f4.eps}}\end{array}
    +\begin{array}{c}\mbox{\epsfig{file=f2.eps}}\end{array}
    \begin{array}{c}\mbox{\epsfig{file=f5.eps}}\end{array}\\
    &\ -\begin{array}{c}\mbox{\epsfig{file=f1.eps}}\end{array}
    \begin{array}{c}\mbox{\epsfig{file=f3.eps}}\end{array}
    \begin{array}{c}\mbox{\epsfig{file=f5.eps}}\end{array}.
\end{array}
\eeq

We can check that our result satisfies (\ref{eq:O}):
\beq
\begin{array}{rl}
m(S\otimes \id)\Delta(\Gamma)&=
    m(S\otimes \id)\left(
    \begin{array}{c}\mbox{\epsfig{file=f7.eps}}\end{array}
    \otimes e
    +e\otimes
    \begin{array}{c}\mbox{\epsfig{file=f7.eps}}\end{array}\right.\\
    &\ \ +\left.
    \begin{array}{c}\mbox{\epsfig{file=f1.eps}}\end{array}
    \otimes
    \begin{array}{c}\mbox{\epsfig{file=f4.eps}}\end{array}
    +\begin{array}{c}\mbox{\epsfig{file=f2.eps}}\end{array}
    \otimes
    \begin{array}{c}\mbox{\epsfig{file=f5.eps}}\end{array}
    \begin{array}{c}\mbox{\epsfig{file=f7.eps}}\end{array}
    \right)\\
    &\ \ 
    +\begin{array}{c}\mbox{\epsfig{file=f7.eps}}\end{array}
    +S\left(
    \begin{array}{c}\mbox{\epsfig{file=f1.eps}}\end{array}
    \right)
    \begin{array}{c}\mbox{\epsfig{file=f4.eps}}\end{array}\\
    &\ \ +S\left(
    \begin{array}{c}\mbox{\epsfig{file=f2.eps}}\end{array}
    \right)
    \begin{array}{c}\mbox{\epsfig{file=f5.eps}}\end{array}\\
    &=
    -\begin{array}{c}\mbox{\epsfig{file=f7.eps}}\end{array}
    +\begin{array}{c}\mbox{\epsfig{file=f1.eps}}\end{array}
    \begin{array}{c}\mbox{\epsfig{file=f4.eps}}\end{array}
    +\begin{array}{c}\mbox{\epsfig{file=f2.eps}}\end{array}
    \begin{array}{c}\mbox{\epsfig{file=f5.eps}}\end{array}\\
    &\ \ -\begin{array}{c}\mbox{\epsfig{file=f1.eps}}\end{array}
    \begin{array}{c}\mbox{\epsfig{file=f3.eps}}\end{array}
    \begin{array}{c}\mbox{\epsfig{file=f5.eps}}\end{array}
    +\begin{array}{c}\mbox{\epsfig{file=f7.eps}}\end{array}
    -\begin{array}{c}\mbox{\epsfig{file=f1.eps}}\end{array}
    \begin{array}{c}\mbox{\epsfig{file=f4.eps}}\end{array}\\
    &\ \ -\begin{array}{c}\mbox{\epsfig{file=f2.eps}}\end{array}
    \begin{array}{c}\mbox{\epsfig{file=f5.eps}}\end{array}
    +\begin{array}{c}\mbox{\epsfig{file=f1.eps}}\end{array}
    \begin{array}{c}\mbox{\epsfig{file=f3.eps}}\end{array}
    \begin{array}{c}\mbox{\epsfig{file=f5.eps}}\end{array}\\
    &=0.
\end{array}
\eeq
where in the third line we used $S(e)=e$. Similarly for (\ref{eq:Or}).

{\bf Example 2.}  On the second example (\ref{eq:example2}), the
antipode is
\beq
S(\Gamma)=-\Gamma-S(\gamma_{1})\cdot\Gamma/\gamma_{1}-
    S(\gamma_{2})\cdot\Gamma/\gamma_{2}
    -S(\gamma_1\cdot\gamma_2)\cdot\Gamma/{(\gamma_1\cdot\gamma_2)}.
\eeq
Both $\gamma_{1}$ and $\gamma_{2}$ are primitive, while
$S(\gamma_1\cdot\gamma_2)=S(\gamma_1)\cdot S(\gamma_2)$, so
\beq
\begin{array}{rl}
S(\Gamma)=&-\Gamma+\gamma_{1}\cdot\Gamma/\gamma_{1}
        +\gamma_{2}\cdot\Gamma/\gamma_{2}
    -(\gamma_1\cdot\gamma_2)\cdot\Gamma/{(\gamma_1\cdot\gamma_2)}\\
 =&
-\begin{array}{c}\mbox{\epsfig{file=ex20.eps}}\end{array}
+\begin{array}{c}\mbox{\epsfig{file=ex23.eps}}\end{array}
\begin{array}{c}\mbox{\epsfig{file=ex21.eps}}\end{array}
+
\begin{array}{c}\mbox{\epsfig{file=ex24.eps}}\end{array}
\begin{array}{c}\mbox{\epsfig{file=ex22.eps}}\end{array}\\
&-
\begin{array}{c}\mbox{\epsfig{file=ex23.eps}}\end{array}
\begin{array}{c}\mbox{\epsfig{file=ex24.eps}}\end{array}
\begin{array}{c}\mbox{\epsfig{file=ex25.eps}}\end{array}.
\end{array}
\eeq

}
\end{quote}

With the above five operations, partitioned spin foams are a Hopf
algebra.  We should emphasize that the same 2-complex with two
different partitionings is two different partitioned spin foams, and
therefore two different elements of the algebra.

We now wish to use the antipode as an extension of the renormalization
group equation appropriate for spin foams.  To do so, we will first
need to consider how block transformations can be incorporated in an
algebra of spin foam weights.

Recall that the spin foam 2-complexes are a convenient pictorial
representation of the weights that contribute in the partition
function (\ref{eq:Z}).  Since they carry the physical information, it
is on these weights that the block transformations apply.  The
labelled partitioned spin foam 2-complexes are in one-to-one
correspondence with parenthesized spin foam weights (of the same spin
foam model).  Therefore, parenthesized weights also form a Hopf
algebra.  We detail this in the next section.

\section{The Hopf algebra of parenthesized spin\\
foam weights}

As we saw in section \ref{weights}, partitioned spin foam lattices 
correspond to
parenthesized spin foam weights.  
Let $W$ be the vector space of parenthesized spin foam weights
over ${\bf C}$.  The following are the Hopf algebra operations on $W$:

$\bullet$ {\bf Multiplication:} Multiplication
$m:W\otimes W\rightarrow W$ is simply the
multiplication of the two spin foam
weight functions, resulting in a new one with overall
nesting structure $()()$:
\beq
m(\omega_{\Gamma_{1}}\otimes\omega_{\Gamma_{2}})
=\omega_{\Gamma_{1}} \cdot\omega_{\Gamma_{2}}
=(\omega_{\Gamma_{1}})(\omega_{\Gamma_{2}}).
\eeq
We call $\omega_{e}$ the weight of the empty spin foam $e$.  Then,
$\omega_{\Gamma}\cdot\omega_{e}=\omega_{e}\cdot\omega_{\Gamma}
=\omega_{\Gamma}$. 

$\bullet$ {\bf Unit:} The unit annihilates weights:
\beq
\epsilon(\omega_{\Gamma})=\left\{\begin{array}{l}
0\qquad\omega_{\Gamma}\neq \omega_{e},\\
1\qquad\omega_{\Gamma}= \omega_{e}.
\end{array}
\right.
\eeq

$\bullet$ {\bf Counit:} The {\em counit} creates weights:
\beq
\coe(c)=c\omega_{e}\qquad c\in{\bf C}.
\eeq

$\bullet$ {\bf Coproduct:} The coproduct, $\Delta:W\rightarrow W\otimes 
W$,
splits the original weight into a sum of the
weights of the subfoams paired with the weight of their remainder:
\beq
\Delta(\omega_{\Gamma})=\omega_{\Gamma}\otimes\omega_{e}
+\omega_{e}\otimes\omega_{\Gamma}
+\sum_{\gamma} \omega_{\gamma}\otimes\omega_{\Gamma/\gamma}.
\eeq
If $\omega_{\gamma_{p}}$ is the weight of a primitive spin foam 
$\gamma_{p}$, then
\beq
\Delta(\omega_{\gamma_{p}})=
\omega_{\gamma_{p}}\otimes e+e\otimes \omega_{\gamma_{p}},
\eeq
while $\Delta(\omega_{e})=\omega_{e}\otimes\omega_{e}$. 

$\bullet$ {\bf Antipode:} 
The antipode is defined iteratively as
\beq
\begin{array}{rl}
S(\omega_{\Gamma})&=-\omega_{\Gamma}-\sum_{\gamma}
S\left(\omega_{\gamma}\right)\cdot \omega_{\Gamma/\gamma},\\
S(\omega_{\gamma_{p}})&=-\omega_{\gamma_{p}},
\end{array}
\label{eq:Somega}
\eeq
which stops at the weight of a primitive spin foam. 
We have $S(\omega_{\Gamma_{1}}\cdot\omega_{\Gamma_{2}})=
S(\omega_{\Gamma_{1}}) S(\omega_{\Gamma_{2}})$, and $S(\omega_e)=e$.

$S$ above is an antipode in $W$ since it satisfies
\beq
m(S\otimes\id)\Delta(\omega_{\Gamma})=
m(\id\otimes S)\Delta(\omega_{\Gamma})=\coe(\omega_{\Gamma}),
\label{eq:Oomega}
\eeq
for all $\omega_{\Gamma}\in W$.

\begin{quote}
{\small
{\bf Example.}  Again, using the spin foam example
(\ref{eq:example}), which has weight
\beq
\omega_\Gamma=\left(\left(\left(
    A_1 A_2 A_3 d_k\right) A_4 A_5 d_l\right)
    A_6 d_i d_j d_m d_n\right),
\label{eq:exomega}
\eeq
we find 
\beq
\begin{array}{rl}
\Delta(\omega_\Gamma)=&\omega_\Gamma\otimes e +e \otimes \omega_\Gamma
    +\left(A_1  A_2 A_3 d_k\right)\otimes
    \left(\left( A_4 A_5 d_l\right)A_6 d_i d_j d_m d_n\right)\\
    &+\left(\left(A_1  A_2 A_3 d_k\right)A_4 A_5 d_l\right)
    \otimes
    \left(A_6 d_i d_j d_m d_n\right).
\end{array}
\eeq
For the antipode, we get
\beq
S(\omega_\Gamma)=-\omega_\Gamma-S(\omega_{\gamma_1})
\cdot\omega_{\Gamma/\gamma_{1}}
    -S(\omega_{\gamma_2})\cdot\omega_{\Gamma/\gamma_{2}}.
    \label{eq:exS}
\eeq
The subfoam
$\gamma_1$ is primitive and therefore $S(\omega_{\gamma_1})=
-\omega_{\gamma_1}$, while for $\omega_{\gamma_2}$ we get
\beq
\begin{array}{rl}
S(\omega_{\gamma_2})&=
    
-\omega_{\gamma_2}-S(\omega_{\gamma_1})\cdot\omega_{\gamma_2/\gamma_1}\\
    &=-\left(\left( A_1  A_2 A_3 d_k\right) A_4 A_5 d_l\right)
    +\left(A_1  A_2 A_3 d_k\right)\left(A_4 A_5 d_l\right).
\end{array}
\eeq
Hence, (\ref{eq:exS}) gives
\beq
\begin{array}{rl}
S(\omega_\Gamma)=&-\omega_\Gamma
        +\left( A_1  A_2 A_3 d_k\right)
    \left(\left(A_4 A_5 d_l\right)A_6 d_i d_j d_m d_n\right)\\
    &+\left(\left( A_1  A_2 A_3 d_k\right)A_4 A_5 d_l\right)
        \left(A_6 d_i d_j d_m d_n\right)\\
    &-\left( A_1  A_2 A_3 d_k\right)
        \left(A_4 A_5 d_l\right)
        \left(A_6 d_i d_j d_m d_n\right).
\end{array}
\label{eq:exSomega}
\eeq
Note that each term in $S(\omega_\Gamma)$ is the weight of the same
{\em unpartitioned} spin foam, but bracketed by different choice of
partitions of the 2-complex.  These are all different elements in
$W$.  Also, note that, in the sense of (\ref{eq:Oomega}), the weight
(\ref{eq:exSomega}) is the inverse of $\omega_\Gamma$ in
(\ref{eq:exomega}).
}
\end{quote}


\section{Block transformations as equivalence relations in the spin
foam algebra}
\label{equiv}

Consider the block transformation
\[
\Gamma =\begin{array}{c}\mbox{\epsfig{file=gamma.eps}}\end{array}
\qquad\longrightarrow \qquad
\Gamma'=\begin{array}{c}\mbox{\epsfig{file=gammap.eps}}\end{array},
\]
which we obtain, for example, by summing over all possible values of
the labels on the edges we eliminated.  
For the corresponding weights, this is a
transformation $\omega_{\Gamma}\rightarrow\omega_{\Gamma'}$.  Now,
$\omega_{\Gamma}$ and $\omega_{\Gamma'}$ are two {\em different} 
functions on
two {\em different} sets of weights.  However, we say that they are {\em
equivalent} if they are related by the renormalization group
equation.
That is, when they both describe the same physical system but at 
different scales. 

The renormalization group transformation is an equivalence relation
\beq
\left(\Gamma',\omega_{\Gamma'}\right):= 
\mbox{RG}\left(\Gamma, \omega_{\Gamma}\right)
\sim\left(\Gamma, \omega_{\Gamma}\right),
\eeq
namely, the coarse-grained lattice and its weight is physically 
equivalent to the original one if
the partition function of the coarse-grained lattice has a value 
appropriately close to the original one, that is,
\beq
\mbox{eval}\ Z(\omega_{\Gamma'})=\mbox{eval}\ 
Z(\omega_{\Gamma})+\mbox{corrections}.
\label{eq:ZRGequiv}
\eeq

In real-space renormalization, $(\Gamma',\omega_{\Gamma'})$ are 
obtained in the following way:
\begin{enumerate}
    \item 
    Subdivide $\Gamma$ into nested/disjoint sublattices 
    $\{\gamma_{i}\}$ (partition $\Gamma$).
    \item
    Repeatedly apply a {\em block transformation} $R(\gamma_{i})$ to 
    the sublattices of $\Gamma$.
\end{enumerate}
Following this analogy, a block transformation $R$ for a spin foam 
should have the following properties:
\begin{enumerate}
    \item
    Acting on a subfoam $\gamma$ of $\Gamma$, it produces a new 
    subfoam $R(\gamma)$ such that $\Gamma/\gamma= 
    \Gamma/R(\gamma)$ (i.e.\ it applies locally to the subfoam).
    \item 
    {\bf Exact block transformation:} $R$ acting on the corresponding 
    weight $\omega_{\gamma}$ produces a new weight 
    $R(\omega_{\gamma})$ on $R(\gamma)$ which is equivalent to the old 
    one.  Equivalence has the same meaning as in eq.\ 
    (\ref{eq:ZRGequiv}), namely, 
    eval$\ Z(R(\omega_{\gamma}))=$eval$\ Z(\omega_{\gamma})$. 
    
    A familiar example of an exact block transformation is summing 
    over all values of the degrees of freedom in the interior of 
    $\gamma$, i.e.\
    \beq
    \begin{array}{rl}
    R(\gamma)&=\partial\gamma,\\
    R(\omega_{\gamma})&=\sum_{\mbox{\footnotesize internal labels}} 
    \omega_{\gamma}.
    \end{array}
    \label{eq:exactR}
    \eeq
    
    {\bf General block transformation:}  $R$ acting on the 
    sublattices $\{(\gamma_{i},\omega_{\gamma_{i}})\}$ in the partition 
    of $\Gamma$, produces a coarse-grained spin foam 
    $(\Gamma',\omega_{\Gamma'})$ which 
    is equivalent to $(\Gamma,\omega_{\Gamma})$ according to 
    (\ref{eq:ZRGequiv}).  It is not the case anymore that 
    $R(\omega_{\gamma})$ is equivalent to $\omega_{\gamma}$.  The 
    equivalence holds only for the final product, 
    $(\Gamma',\omega_{\Gamma'})$.
    
    Examples of such more general block transformations are 
    decimation, truncation, etc.  Decimation of a sublattice produces 
    a new one which is not in any sense equivalent to the original 
    sublattice, but after decimation has been applied to every 
    sublattice in the partition of $\Gamma$, the final product will 
    be equivalent to $\Gamma$ (when decimation works, naturally).
\end{enumerate}

We want to incorporate the physical equivalence that the 
renormalization group is based on as an equivalence relation in the 
Hopf algebra of partitioned spin foams and, in this article, we want 
to do this for real-space renormalization.  The idea is the following.
The renormalization group is the equivalence
\beq
\Gamma'-\Gamma\stackrel{\mbox{\footnotesize RG}}{\sim} 0.
\eeq
This becomes an equivalence relation in the algebra when we write it as 
\beq
\Gamma'-\Gamma=m\left(S_{R}\otimes\id\right)\Delta(\Gamma)
\stackrel{\mbox{\footnotesize RG}}{\sim}0,
\eeq
where $S_{R}$ is a modification of the antipode in which we have 
inserted the transformation operation $R$ so that it block 
transforms the sublattices of $\Gamma$ as they appear on the left of 
the tensor product in $\Delta(\Gamma)$.  

Of course, the antipode for a Hopf algebra is unique, so any
modification of it will give an operation which is not an antipode. 
The idea is that if $R$ and the modification of the antipode are
chosen so that they contain the physical equivalence in the
renormalization group, then $S_{R}$ should be an antipode in the
algebra under the equivalence relation that the RG defines:
\beq
m\left(S_{R}\otimes\id\right)\Delta(\Gamma)
\stackrel{\mbox{\footnotesize RG}}{\sim}0.
\eeq

All this is easier and clear to see if we use an exact block 
transformation $R$.  In the next section we use the $R$ defined in 
(\ref{eq:exactR}) and an appropriate modified antipode on spin foam 
examples.  The general case (general $R$) is discussed in section 
\ref{section9}.


\section{Using the antipode to perform local scale transformations: 
exact transformations}

We will
obtain local scale transformations of a spin foam by inserting the
equivalence relation $R$ as defined in (\ref{eq:exactR}) simultaneously
in the antipode (\ref{eq:S}) in $V$ and (\ref{eq:Somega}) in $W$.
In effect, we will define a new antipode for $V$ and $W$, which
performs local scale transformations on the spin foams.  We call
this modified antipode a {\em shrinking antipode}.  It is
equivalent to (\ref{eq:S}) and (\ref{eq:Somega}) under
the equivalence relation $R$ (and therefore it is an 
antipode in the algebra only under this equivalence relation).

The action of an exact block transformation is significantly
different to that of an approximate one, and it turns out that
different modifications of the antipode are appropriate in each
of these cases. In this section, we define a shrinking antipode 
appropriate for exact block transformations.  

A possible definition of an exact
block transformation operation $R$ on $V$ is the following:
\beq
R(\gamma)=\begin{array}{l}
\mbox{in every connected component of }\gamma\mbox{ all 
faces are eliminated  }\\
\mbox{and the vertices are shrunk to a single vertex } 
\begin{array}{c}\mbox{\epsfig{file=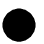}}\end{array}.
    \end{array}
    \label{eq:Gre}
\eeq
Then, $R$ on $\omega_{\gamma}$ sums over all possible values of the 
labels on the faces of $\gamma$:
\beq
R(\omega_\gamma)={{{\sum}}_{ j_f, f\in\gamma}}
\ \omega_\gamma,
\label{eq:Ore}
\eeq
In effect,
$R(\omega_\gamma)$ reduces $\omega_\gamma$ to a product of
amplitudes, one for each connected component of $\gamma$.
Note that, for $\Gamma$ itself, $R(\Gamma)$ eliminates all faces 
except those who belong to the in and out spin networks (see example 
below).

Given such an exact renormalization scheme $R$, we modify $S$ in
(\ref{eq:S}) and (\ref{eq:Somega}) by inserting $R$ as follows:
\beq
\begin{array}{rl}
\sre(\Gamma)&=-R(\Gamma)-\sum_{\gamma}\sre(\gamma)\cdot\Gamma/\gamma,\\
\sre(\gamma_p)&=-R(\gamma_p)\\
\sre(e)&=e.
\end{array}
\label{eq:SRexact}
\eeq 
On the corresponding weights we have:
\beq
\begin{array}{rl}
\sre(\omega_\Gamma)&=-R(\omega_\Gamma)-
\sum_{\gamma} \sre(\omega_\gamma) \cdot \omega_{\Gamma/\gamma},\\
\sre(\omega_{\gamma_p})&=-R(\omega_{\gamma_p})\\
\sre(\omega_{e})&=\omega_{e}. 
\end{array}
\label{eq:SRexactw}
\eeq
We call  $\sre$
the {\em exact shrinking antipode}.

\begin{quote}
{\small
{\bf Example:}
Consider again the spin foam (\ref{eq:example}),
\[
\Gamma=\begin{array}{c}\mbox{\epsfig{file=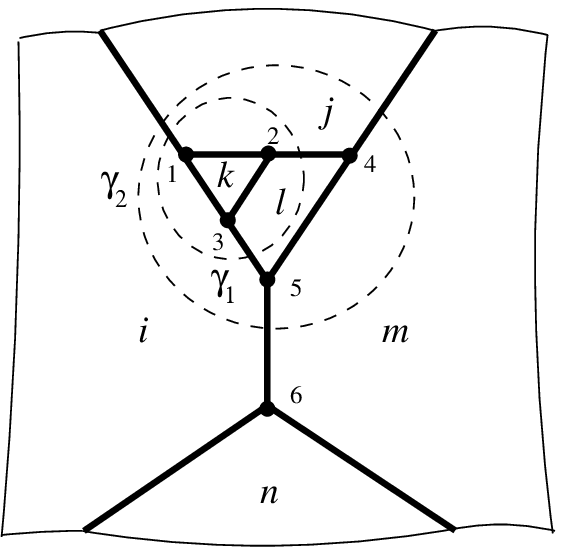}}
\end{array}
\]
with weight
\beq
\omega_\Gamma=\left(\left(\left(A_1 A_2 A_3 d_k\right)
A_4 A_5 d_l\right)
A_6 d_i d_j d_m d_n\right).
\eeq

Then, from (\ref{eq:SRexact}) and (\ref{eq:SRexactw}),  the  shrinking antipode of $\Gamma$
and $\omega_{\Gamma}$ is
\beq
\begin{array}{rl}
\sre(\Gamma)&=-R(\Gamma)-\sre(\gamma_1)\cdot\Gamma/\gamma_1
    -\sre(\gamma_2)\cdot\Gamma/\gamma_2,\\
&\\
\sre(\omega_\Gamma)&=-R(\omega_\Gamma)-\sre(\omega_{\gamma_1})
    \omega_{\Gamma/\gamma_1}
    -\sre(\omega_{\gamma_2})\omega_{\Gamma/{\gamma_2}}.
\end{array}
\label{eq:exg}
\eeq

For $\gamma_1$, we have
\beq
\begin{array}{rl}
\sre(\gamma_1)=&\begin{array}{c}\mbox{\epsfig{file=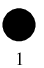}}\end{array}\\

\sre(\omega_{\gamma_1})=&\sum_k \left(A_1 A_2 A_3 d_k\right)
=N A'_1,
\end{array}
\eeq
where $N=\sum_i d_i$ for the group $G$ and $A'_1=A_1 A_2 A_3$.

For $\gamma_2$ we get
\beq
\begin{array}{rl}
\sre(\gamma_2)&= -R(\gamma_2)-\sre(\gamma_1)\cdot\gamma_2/\gamma_1\\
    &\\
    &=-\begin{array}{c}\mbox{\epsfig{file=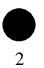}}\end{array}
    +\begin{array}{c}\mbox{\epsfig{file=dot1.eps}}\end{array}
    \begin{array}{c}\mbox{\epsfig{file=f3.eps}}\end{array},\\
\sre(\omega_{\gamma_2})&=-R(\omega_{\gamma_2})-\sre(\omega_{\gamma_1})
    \omega_{\gamma_2/\gamma_1}\\ 
    &\\
    &=-\sum_{k,l} \left(\left( A_1 A_2 A_3 d_k\right)
        A_4 A_5 d_l\right)-R(\omega_{\gamma_1})
                \omega_{\gamma_2/\gamma_1}\\
    &\\
    &=-N^2 A_2'+N A_1'\left( A_4 A_5 d_l\right),
\end{array}
\eeq
where $A_2'=A_1 A_{2} A_{3} A_{4} A_5$.

Finally,  $R(\Gamma)$ is the spin foam
\beq
R(\Gamma)=\begin{array}{c}\mbox{\epsfig{file=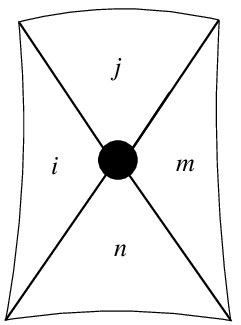}}\end{array}
\eeq
with weight
\beq
R(\omega_\Gamma)={{\sum}}_{k,l}\ \omega_\Gamma= N^2 A'\  d_i d_j d_m d_n,
\eeq
where $A'=A_1 A_{2} A_{3} A_{4} A_5 A_6$.

Plugging the results back to equations (\ref{eq:exg}),
we get the answer
\beq
\begin{array}{rl}
\sre(\Gamma)=&
-\begin{array}{c}\mbox{\epsfig{file=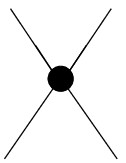}}\end{array}
+\begin{array}{c}\mbox{\epsfig{file=dot1.eps}}\end{array}
\begin{array}{c}\mbox{\epsfig{file=f4.eps}}\end{array}
+\begin{array}{c}\mbox{\epsfig{file=dot2.eps}}\end{array}
\begin{array}{c}\mbox{\epsfig{file=f5.eps}}\end{array}\\
&-\begin{array}{c}\mbox{\epsfig{file=dot1.eps}}\end{array}
\begin{array}{c}\mbox{\epsfig{file=f3.eps}}\end{array}
\begin{array}{c}\mbox{\epsfig{file=f5.eps}}\end{array}
\end{array}
\label{eq:Gsreexample}
\eeq
for the 2-complex, and
\beq
\begin{array}{rl}
\sre(\omega_\Gamma)=
    &-N^2 A' d_i d_j d_m d_n
    +N A_1'\left(\left( A_4 A_5 d_l\right) A_6 d_i d_j d_m d_n\right)\\
    &\\
    &+N^2 A_2'\left(A_6 d_i d_j d_m d_n\right)
    -N A_1'\left(A_4 A_5 d_l\right)\left(A_6 d_i d_j d_m d_n\right)
\end{array}
\label{eq:Osreexample}
\eeq
for its weight.
}
\end{quote}

The {\em fully shrunk} spin foam $\Gamma'$
(all faces eliminated except those who belong to the in and out spin 
networks)
is given by
\beq
\begin{array}{rl}
\Gamma'&=\Gamma-m(\sre\otimes\id)\Delta(\Gamma)\\
\omega_{\Gamma'}&=\omega_\Gamma-m(\sre\otimes\id)\Delta(\omega_\Gamma).
\end{array}
\label{eq:gprime}
\eeq
Note that,
for the exact block transformation we defined in (\ref{eq:Gre}) and 
(\ref{eq:Ore}), $\Gamma'$ is the same as
$R(\Gamma)$.  However, 
this is only true for this particular $R$.  In general, $R(\Gamma)$ is 
not the renormalized spin foam.  We discuss this in detail in the next 
section.

Also note that the fully shrunk spin foam is a single amplitude, and
the same as the evaluation of the spin foam.

\begin{quote}
    \small{
    {\bf Example:}
To fully coarse-grain our example (\ref{eq:example}) 
we use equation (\ref{eq:gprime}) to get:
\beq
\begin{array}{rl}
\Gamma'=&\Gamma-m(\sre\otimes\id)\left(\Gamma\otimes e+e\otimes\Gamma   
+\gamma_1\otimes\Gamma/\gamma_1+\gamma_2\otimes\Gamma/\gamma_2\right)\\
    &\\
    =&\Gamma-\sre(\Gamma)-\Gamma-\sre(\gamma_1)\cdot\Gamma/\gamma_1
        -\sre(\gamma_2)\cdot\Gamma/\gamma_2\\
    &\\
    =&\begin{array}{c}\mbox{\epsfig{file=f7.eps}}\end{array}
    +\begin{array}{c}\mbox{\epsfig{file=f6.eps}}\end{array}
    -\begin{array}{c}\mbox{\epsfig{file=dot1.eps}}\end{array}
    \begin{array}{c}\mbox{\epsfig{file=f4.eps}}\end{array}
    -\begin{array}{c}\mbox{\epsfig{file=dot2.eps}}\end{array}
    \begin{array}{c}\mbox{\epsfig{file=f5.eps}}\end{array}\\
    &+\begin{array}{c}\mbox{\epsfig{file=dot1.eps}}\end{array}
    \begin{array}{c}\mbox{\epsfig{file=f3.eps}}\end{array}
    \begin{array}{c}\mbox{\epsfig{file=f5.eps}}\end{array}
    -\begin{array}{c}\mbox{\epsfig{file=f7.eps}}\end{array}
    +\begin{array}{c}\mbox{\epsfig{file=f4.eps}}\end{array}\\
    &+\begin{array}{c}\mbox{\epsfig{file=dot2.eps}}\end{array}
    \begin{array}{c}\mbox{\epsfig{file=f5.eps}}\end{array}
    -\begin{array}{c}\mbox{\epsfig{file=dot1.eps}}\end{array}
    \begin{array}{c}\mbox{\epsfig{file=f3.eps}}\end{array}
    \begin{array}{c}\mbox{\epsfig{file=f5.eps}}\end{array}\\
    =&\begin{array}{c}\mbox{\epsfig{file=f6.eps}}\end{array}.
\end{array}
\eeq
}
\end{quote}

The shrinking antipode acting on a spin foam $\Gamma$ produces a new
foam which is a sum over all spin foams that can be obtained by {\em
local} block transformations of $\Gamma$.  That is, the different
terms in the equations (\ref{eq:Gsreexample}) and
(\ref{eq:Osreexample}) are all the original spin foam $\Gamma$ with
different parts of it coarse-grained.  The coarse-graining is
inhomogeneous, resulting in spin foams whose labels carry couplings at
different scales (coarse-grained vertex amplitudes correspond to
different multiples of $l_{pl}$.  The sign of each term ensures that
this sum over all possible local coarse-grainings is itself equivalent
to the original spin foam.

This, of course, is not how coarse-graining is done in statistical
physics for a usual lattice system, as there is no need and no
advantage in this proliferation of terms (although it gives  some
insight into the renormalization group, as we will see in section
\ref{RG}).  But recall that a main issue in developing renormalization
group methods for spin foams is that the lattices are
background-independent and irregular.  A global block-transformation,
as in standard real space renormalization, is not only difficult to
implement, but it is not even clear if it is physically meaningful. 
Local ones are, and it is those that we use here.  We will discuss
this in some more detail in the next section, where we compare $S_{R}$
to the usual renormalization group.


\section{$\sre$ vs the renormalization group equation}
\label{RG}

In this section, we compare block transformations carried out using
the exact shrinking antipode $\sre$ and the standard renormalization 
group
equation.  This is best done by example, and we next calculate $\sre$
for a familiar {\em regular} lattice.

Consider the square lattice
\beq
\Gamma=\begin{array}{c}\mbox{\epsfig{file=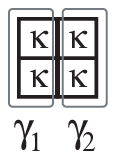}}\end{array}
\eeq
e.g., a square lattice in $Z_{2}$ lattice 
gauge theory
with spins $\pm 1$ on the edges and couplings $\kappa$ on the
plaquettes.  It is partitioned as marked, that is, into sublattices 
and remainders
\beq
\begin{array}{lcl}
    \gamma_{1}=\begin{array}{c}\mbox{\epsfig{file=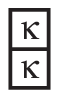}}\end{array},
     &\qquad &
     
\Gamma/\gamma_{1}=\begin{array}{c}\mbox{\epsfig{file=sq1.eps}}\end{array}
\\
    
    \gamma_{2}=\begin{array}{c}\mbox{\epsfig{file=sq1.eps}}\end{array},
     &\qquad &
     
\Gamma/\gamma_{2}=\begin{array}{c}\mbox{\epsfig{file=sq1.eps}}\end{array}.

\\
\end{array}
\eeq
We have made the obvious choices of what a sublattice should be in 
this case so that $\omega_{\gamma}\cdot\omega_{\Gamma/\gamma}= 
\omega_{\Gamma}$.  Note that $\gamma_{1}\neq \gamma_{2}$ since they 
have different labels on the edges. 

With this partition, we calculate the coproduct of $\Gamma$ to be 
\beq
\Delta(\Gamma)=\begin{array}{c}\mbox{\epsfig{file=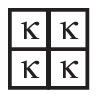}}\end{array}
\otimes e
+
e\otimes
\begin{array}{c}\mbox{\epsfig{file=sq.eps}}\end{array}
+
\begin{array}{c}\mbox{\epsfig{file=sq1.eps}}\end{array}
\otimes
\begin{array}{c}\mbox{\epsfig{file=sq1.eps}}\end{array}
+
\begin{array}{c}\mbox{\epsfig{file=sq1.eps}}\end{array}
\otimes
\begin{array}{c}\mbox{\epsfig{file=sq1.eps}}\end{array}.
\eeq

We can use a standard exact block transformation $R$ which removes all 
internal edges in a sublattice:
\beq
R(\gamma)=\partial\gamma,
\eeq
by summing over all values of the labels on these edges:
\beq
R(\omega_{\gamma})=
\sum_{\mbox{\footnotesize labels on internal edges of }\gamma}
\omega_{\gamma}.
\eeq
With this $R$, we calculate the exact shrinking antipode of eq.\ 
(\ref{eq:SRexact}) to be
\beq
\begin{array}{rl}
    \sre(\Gamma)&=
    -R\left(
    \begin{array}{c}\mbox{\epsfig{file=sq.eps}}\end{array}\right)
    -\sre\left(
    \begin{array}{c}\mbox{\epsfig{file=sq1.eps}}\end{array}
\right)\cdot
\begin{array}{c}\mbox{\epsfig{file=sq1.eps}}\end{array}
    -\sre\left(
    \begin{array}{c}\mbox{\epsfig{file=sq1.eps}}\end{array}
\right)\cdot
\begin{array}{c}\mbox{\epsfig{file=sq1.eps}}\end{array}
    \\  
&\\
&=
-\begin{array}{c}\mbox{\epsfig{file=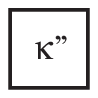}}\end{array}
+\begin{array}{c}\mbox{\epsfig{file=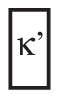}}\end{array}\cdot
\begin{array}{c}\mbox{\epsfig{file=sq1.eps}}\end{array}
+\begin{array}{c}\mbox{\epsfig{file=sq2.eps}}\end{array}\cdot
\begin{array}{c}\mbox{\epsfig{file=sq1.eps}}\end{array}
\\
&\\
&=
-\begin{array}{c}\mbox{\epsfig{file=sq3.eps}}\end{array}
+\begin{array}{c}\mbox{\epsfig{file=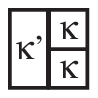}}\end{array}
+\begin{array}{c}\mbox{\epsfig{file=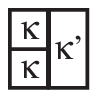}}\end{array}.
\\
\label{eq:sresquare}
\end{array}
\eeq
In the last line above, we have carried out the multiplications of the 
weights and rejoined the sublattices to match the labels on the 
edges.  Again, in the first two lines, the last two terms are 
different, they have different labels on the edges.

In comparison, the standard renormalization group equation on $\Gamma$ 
gives 
\beq
\mbox{RG}\left(\begin{array}{c}\mbox{\epsfig{file=sq.eps}}\end{array}
\right)=
\begin{array}{c}\mbox{\epsfig{file=sq3.eps}}\end{array}.
\eeq
We can understand the RG transformation as a special case of 
$\sre$ in which only the homogeneous terms appear.  In this
example, homogeneous terms are lattices containing only square 
plaquettes, all carrying the same coupling.  In (\ref{eq:sresquare}) 
we have only one such term, the first, but clearly, if we had started 
with a larger lattice, all the possible homogeneous lattices that can 
be obtained by coarse-graining $\Gamma$ would appear.  

As we saw in section \ref{equiv}, the
correspondence between RG and $\sre$ can be made precise as 
follows.  Let us call $\Gamma^{\prime}$ the result of the RG 
transformation:
\beq
\Gamma^{\prime}:={\mbox{RG}}\left(\Gamma\right).
\eeq
$\Gamma^{\prime}$ is physically equivalent to $\Gamma$, since
they are related by the renormalization group transformation:
\beq
\Gamma-\Gamma^{\prime}\stackrel{\mbox{\footnotesize RG}}{\sim}0.
\eeq
We may rewrite this as 
\beq
\Gamma-\Gamma^{\prime}=m\left(\sre\otimes\id\right)\Delta(\Gamma),
\eeq
since $m\left(\sre\otimes\id\right)\Delta(\Gamma)
\stackrel{\mbox{\footnotesize RG}}{\sim}0$.  

We can check this for our example. 
We have 
\beq
\begin{array}{rl}
m\left(\sre\otimes\id\right)\Delta(\Gamma)
  =& -\begin{array}{c}\mbox{\epsfig{file=sq3.eps}}\end{array} 
  +\begin{array}{c}\mbox{\epsfig{file=sq4.eps}}\end{array}
  +\begin{array}{c}\mbox{\epsfig{file=sq5.eps}}\end{array}\\
  &\\
  &
  +\begin{array}{c}\mbox{\epsfig{file=sq.eps}}\end{array}
  -\begin{array}{c}\mbox{\epsfig{file=sq4.eps}}\end{array}
  -\begin{array}{c}\mbox{\epsfig{file=sq5.eps}}\end{array}\\
  &\\
  =&-\begin{array}{c}\mbox{\epsfig{file=sq3.eps}}\end{array}
  +\begin{array}{c}\mbox{\epsfig{file=sq.eps}}\end{array}.
\end{array}
\eeq
which indeed gives
\beq
\Gamma^{\prime}=\Gamma-m\left(\sre\otimes\id\right)\Delta(\Gamma)
       =\begin{array}{c}\mbox{\epsfig{file=sq3.eps}}\end{array}.
\eeq

In this sense, the standard RG equation is embedded in the shrinking 
antipode.  The antipode is an expansion of $\Gamma$ into all its 
possible coarse-grainings, 
with the signs in the different terms combining to
respect what the RG encodes: 
the physical equivalence of two descriptions of the same 
system at different scales.  
It is very interesting to note that we can revisit the 
statement that ``the renormalization group is not a group because it 
has no inverse'' and propose that it is a Hopf algebra and it has 
an antipode!

One might object that on the practical, calculational side,
there are no advantages 
in coarse-graining using $\sre$ since 
the wanted result $\Gamma^{\prime}$ is one of the terms in $\sre$, 
namely, $R(\Gamma)$, and there is no need to calculate the 
other terms.  However, this is a coincidence of our choice of block 
transformation $R$.  This $R$ is itself an equivalence relation,
$\gamma'\stackrel{R}{\sim}\gamma$ for all sublattices $\gamma$, and 
in particular
\beq
\Gamma'\stackrel{R}{\sim}\Gamma,
\eeq
which leads to the oversimplification $\Gamma'=R(\Gamma)$ as a term 
in the antipode. 

Other choices of $R$, for example, expanding 
$\Gamma$ and keeping only the lowest order terms, requires all the 
terms in $\sre$ to calculate $\Gamma^{\prime}$ (the coarse-grained 
system that is physically equivalent to $\Gamma$).
For such choices of approximate block transformations, alternative 
definitions of the shrinking antipode are also possible.  We discuss this 
in the next section.

Closing this section, it is important to note that the RG equation can
only be applied to systems with a background.  In real space
renormalization we always pick partitions of the lattice into
identical sublattices, and block transform each one so that a RG step
takes us from couplings $\{\kappa\}$ everywhere, to couplings
$\{\kappa^{\prime}\}$ everywhere.  We need the lattice spacing as a
guide everywhere on the lattice, which means we need a background and
we also need to have regular lattices.  Spin foams are background 
independent and highly irregular.
Coarse-graining via the antipode does not require a
global choice of lattice spacing, or regular lattices, and so it
can be applied to spin foams.  It can also be applied to irregular
lattices with a background, where it may provide a useful
calculational tool.  Its strong point is that the antipode is defined
as an iteration, and therefore it is suited for numerical calculations
(as was shown, for example, in \cite{BrKr} for quantum field theory).

\section{$S_{R} $ and general block transformations}
\label{section9}

In the previous two sections, we gave detailed examples of exact 
block transformations of both spin foams and lattice gauge theory, 
using the exact shrinking antipode $\sre$.  The modification of $S$ 
to $\sre$ that we used is the simplest choice that satisfies 
$m(\sre\otimes\id)\Delta(\Gamma)=\coe(\Gamma)$ so that $\sre$ is still 
the antipode in the spin foam algebra.  To check the equivalence of 
$\sre$ to $S$, we had the advantage that an exact $R$ is itself an 
equivalence relation. 

However, it is clear that spin foams are complicated enough models 
that solving them using exact $R$ is unlikely to be practical.  
Giving specific examples of approximate coarse-graining schemes is 
beyond the scope of this article.  In this section, we discuss the 
general features of such a coarse-graining using the Hopf algebra.

The basic idea is again the same.  We will coarse-grain using a 
shrinking antipode $S_{R}$ which is equivalent to the original one 
$S$ under the physical equivalence of the renormalization group 
transformation, namely, 
\[
\Gamma'-\Gamma=m\left(S_{R}\otimes\id\right)\Delta(\Gamma)
\stackrel{\mbox{RG}}{\sim}0
\]
(for non-empty $\Gamma$) where $\Gamma'$ is the renormalized spin 
foam.  Clearly, further conditions on $R$ and $S_{R}$ are also 
required, for example $R[R(\Gamma)]=R(\Gamma)$, and a choice of 
$S_{R}$ that preserves the Hopf algebra structure (associativity, 
coassociativity etc). 

We should emphasize that we have generalized the RG to apply to sums 
over lattices, which is necessary for spin foams.  In spin foam 
models, we commonly use
\beq
{\bf \Gamma}(s_{1},s_{2})=\sum_{\partial\Gamma_{i}=s_{1}\cup 
s_{2}} \Gamma_{i},
\eeq
namely, sums over all spin foams with a given boundary, spin 
networks $s_{1}$ and $s_{2}$.  Since a general element of our algebra 
has the form $\Gamma=\sum c_{i}\Gamma_{i}$, $c\in{\bf C}$, such ${\bf 
\Gamma}(s_{1},s_{2})$ are simply particular elements of the algebra.  
Note that each $\Gamma_{i}$ in the sum is a {\em partitioned} spin foam. 
Also note that, as Kreimer showed in the original version of the Hopf 
algebra, the generators of the algebra are the 1-particle irreducible 
diagrams (here, the straightforward generalization of 1PIs to 
2-complexes).

Also note that equivalence relations other than block transformations 
can be inserted in the algebra operations. In particular, we can use 
the recoupling moves as extra equivalence relations.  For example, in
2-dimensional spin foams, we can use 
\[
R\left(\begin{array}{c}\mbox{\epsfig{file=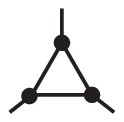}}\end{array}
\right)=
\begin{array}{c}\mbox{\epsfig{file=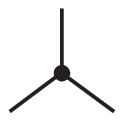}}\end{array},
\]
together with 
\[
M\left(\begin{array}{c}\mbox{\epsfig{file=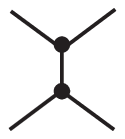}}\end{array}
\right)=
\begin{array}{c}\mbox{\epsfig{file=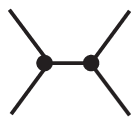}}\end{array},
\]
in $S_{R}$ to keep the number of primitive spin foams small. 

Concluding the general case of $S_{R}$ scale transformations, we will 
also observe that 
in \cite{ReRoFT,DFKR,Pfei}, spin foams are written as a field theory over a group.  
That case is very close in form to the original Kreimer Hopf algebra 
for QFT renormalization, so that is a first place where a general 
$S_{R}$ can be tested. 
Presumably, the form of $S_{R}$ to be used there is 
the same as Kreimer's, namely,
\beq
S_{R}(\Gamma)=-R(\Gamma)-R\left(\sum_{\gamma} 
S_{R}(\gamma)\Gamma/\gamma\right).
\eeq

Further work is required to set up appropriate approximate schemes or
other calculations, such as linearizing near fixed points etc.  We
discuss this further in the Conclusions.

\section{Conclusions}

Our aim in this article was to formulate the problem of finding the 
low-energy limit of spin foam models as a coarse-graining problem, in 
the sense of statistical physics.  Spin foam models, however, differ 
from solid state models in fundamental ways: they model spacetime 
itself.  Therefore, standard renormalization group tools are not 
applicable here.  In particular: 1) there is no background and 
therefore no useful notion of lattice spacing, 2) they are highly 
irregular and 3) they are sums over lattices. 

We proposed that one can deal with all of these features by using a 
generalization of the usual renormalization group.  This is a 
modification of the Kreimer Hopf algebra to real-space renormalization.
The elements of the algebra are sums over partitioned spin foams.  
This method uses the nesting structure of partitioned spin foams 
rather than lattice spacing. 

As we discussed in section 3, there are two parts to coarse-graining.  
The first is to find the right block transformation for the system, 
the second is to repeatedly apply this to the entire lattice.  Here, 
the first part is done using an operation $R$ that applies to 
subfoams.  The second is a combinatorial problem, trivial for regular 
lattices but difficult for irregular ones.  We control it using a 
modification of the algebra antipode in which we have inserted $R$ 
appropriately so that scale transformations are equivalence relations 
in the algebra.

Coarse-graining using the Hopf algebra antipode applies to spin foams
in any dimension.  This is because the Kreimer Hopf algebra, which in
this paper we applied to spin networks, is an algebra of rooted trees. 
These only encode the nesting of subfoams in the spin foam.  This
works in any dimension, one only needs to be careful in the definition
of the subfoams.

We gave explicit expressions for $R$ and the modified antipode in the 
case of exact block transformations.  However, one expects that it is 
approximate schemes that will be most relevant for spin foams and, in 
fact, it is here that the Hopf algebra is expected to be most 
powerful since the antipode is an iteration and thus suited to 
numerical calculations.  We discussed the general features of that 
case in the previous section.  Providing explicit expressions for 
specific spin foam models is an entire research program and beyond 
the scope of this paper. 

The following are some of the obvious directions for further work 
using this algebra:  1) We have generalized the renormalization group 
equation and found that it is embedded in the antipode.  One should 
understand the analogue of renormalization group flows for $S_{R}$.  
2) This can be used to do linearization around fixed points.  In 
particular, one can show that topological quantum field theories are 
fixed points for spin foams and linearize around them to obtain 
near-topological models.  3)  Any local operations on spin foams act 
on subfoams.  The Hopf algebra is then 
an appropriate framework for any local operations, not only scale 
transformations. 

Closing this article, we would like to repeat that the renormalization
group is a general framework and not a recipe that universally applies
to any system.  Further progress should be made by analyzing specific
models.  In this direction, one should keep in mind that experience
from statistical physics teaches us that identifying the correct
models for the systems we are interested in, experimental input is
required.
\section*{Acknowledgments}

I would like to thank Renate Loll and Thomas Thiemann for discussions
and the Albert Einstein Institute, where most of this work was carried
out, for its hospitality.  This work was supported by the EU Network
on ``Discrete Random Geometry'' grant HPRN-CT-1999-00161.


\end{document}